\documentclass[prl,twocolumn,showpacs,preprintnumbers,amsmath,amssymb]{revtex4}
\usepackage{array}
\usepackage{graphicx}
\usepackage{dcolumn}
\usepackage{bm}
\begin{document}

\title{Intrinsic Electric Dipole Moments of Paramagnetic Atoms: Rubidium and Cesium}
\author{H. S. Nataraj$^{1,2}$}\email{nataraj@iiap.res.in} \author{B. K. Sahoo$^3$} \author{B. P. Das$^1$ } \author{D. Mukherjee$^4$}
\affiliation{$^1$ Indian Institute of Astrophysics, Bangalore 560 034, India} 
\affiliation{$^2$ Department of Physics, Mangalore University, Mangalore 574199, India}
\affiliation{$^3$ KVI, University of Groningen, NL-9747 AA Groningen, The Nederlands } 
\affiliation{$^4$ Indian Association for the Cultivation of Sciences, Kolkata 700 032, India}

\date{\today}
\begin{abstract}
The electric dipole moment (EDM) of paramagnetic atoms is sensitive to the intrinsic EDM contribution from that of its constituent electrons and a scalar--pseudo-scalar (S-PS) electron--nucleus interactions. 
The electron EDM and the S-PS EDM contribution to atomic EDM scales as $\approx Z^3$. Thus, the heavy paramagnetic atomic systems will exhibit large enhancement factors. However, the nature of the coupling is so small that it becomes an interest of high precision atomic experiments. In this work we have computed the EDM enhancement factors of the ground states of Rb and Cs due to both the electron EDM and the S-PS EDM using the relativistic coupled-cluster (RCC) theory. The importance of obtaining the precise enhancement factors and the experimental results in deducing a reliable limit on the electron EDM is emphasized. 
\end{abstract}


\maketitle

In the contemporary era of Large Hadron Collider (LHC) in which physicists worldwide are waiting with excitement and hoping to find answers to many outstanding questions haunting the field of particle physics for decades, 
non-accelerator probes like the observation of the electric dipole moment (EDM) of any elementary particle attempting to look for new physics is quite significant. It is indeed complementary to the efforts of LHC,  however, at the low energy regime. In fact, despite the relentless search for a non-zero EDM of elementary particle like the electron for more than 50 years, no conclusive results have been obtained so far. This is quite intriguing and a challenging task for the experimentalists because of the nature of the couplings involved. The EDM of the electron if observed would be extremely small ($ < 1.6 \times 10^{-27} e-cm$ \cite{regan02}) and hence it certainly falls into the category of the highest precision atomic physics experiments. The extensive search for EDMs has resulted in the application of several ingenious experimental techniques including a variety of cooling and trapping techniques, electric and magnetic shielding mechanisms etc. These latest low energy high precision EDM experiments on Rb and Cs atoms are being pursued in different laboratories \cite{weiss, gould, heinzen} and many proposals are still being considered including an experiment to be conducted in the microgravity environments on-board space missions \cite{nasa-edm}. All these proposed EDM experiments are striving hard to improve the sensitivities by 2-3 orders of magnitude higher than the current limits. It would be possible to obtain better limit on the electron EDM (e-EDM) by combining the precise theoretical and experimental results. In this context, we have carried out high precision numerical calculations of the EDM enhancement factors, first in the series, of Rb and Cs which are sensitive to the intrinsic EDM of the electron and the scalar--pseudo-scalar (S-PS) electron-nucleus interactions \cite{pospelov}.

The e-EDM has far reaching consequences ranging from cosmology \cite{huber07} to particle physics \cite{bernreuther91}. The non-zero observation of EDM in any non-degenerate system will be a direct proof of time-reversal violation in nature. Various models of particle physics have predicted different values for the magnitude of EDM of the electron, many of which are close to the current experimental limits unlike the value predicted by the Standard Model which cannot be achieved even in a foreseeable future and thus the observation of non-zero EDM can be helpful in constraining different models of particle physics which predict the size and nature of CP violation which, in turn, has wider implications such as understanding the baryon asymmetry of the universe \cite{bernreuther02}, masses of certain heavy particles, of course, depending on the type of the model considered \cite{bernreuther91}. 
 Thus EDMs undoubtedly will unfold a novel direction for understanding the physics beyond the Standard Model.

We employ the open-shell relativistic coupled-cluster (RCC) theory with single, double and a subset of leading triple excitations in our calculation which is known by the name CCSD(T) method in the literature. The general discussion of the CCSD(T) method is described in \cite{bijaya-pnc} and the references therein. The method employed in particular in calculating the EDM enhancement factors is discussed in detail in \cite{nataraj1,bijaya-sps}. However, for the sake of completeness, we briefly outline the procedure here.

 We first obtain a reference wavefunction ($|\Phi_0\rangle$) for the closed-shell state with $N-1$ electrons, $N$ is the total number of electrons in the system including
a valence electron ($v$), by solving the Dirac-Fock (DF) equations; the Hamiltonian of which is given by,
\begin{eqnarray}
H_0 = \sum_i \{ c \alpha_i \cdot p_i + (\beta_i - 1)m c^2 + V_{\mathcal{N}}(r_i)\} +  \sum_{i<j} V_C(r_{ij})
\label{H0}
\end{eqnarray}
where $c$ is the speed of light in vacuum, $\alpha$ and $\beta$ are the Dirac matrices, $V_{\mathcal{N}}$ is the nuclear potential and $V_C$ is the Coulomb interaction term.

In the framework of the RCC theory, we construct the
exact wavefunction ($|\Psi_v^{(0)}\rangle$) for the corresponding valence electron system as,
\begin{eqnarray}
\vert \Psi_v^{(0)}\rangle = e^{T^{(0)}}\{ 1 + S_v^{(0)}\} \vert\Phi_v\rangle
\label{eqn3}
\end{eqnarray}
where $T^{(0)}$ is the excitation operator for the core (occupied) orbital electrons, $S_v^{(0)}$ corresponds to the excitation operator for the valence and valence-core electrons and $\vert\Phi_v\rangle = a_v^\dagger\vert\Phi_0\rangle $ is the single valence reference state  where $a_v^\dagger$ is the particle creation operator.

Even in the absence of any external field, any paramagnetic atom may have two dominant sources of EDMs; one, the intrinsic e-EDM contribution and the other due to P- and T- violating S-PS electron--nucleus interactions. They are neglected due to their meager influences on determining any physical quantities. However, their effects are considerably larger in heavy atomic systems and can be studied to probe some subtle effects. The intrinsic e-EDM contribution is purely quantum field theoretical in nature and will be non-vanishing only in the relativistic treatment of the problem \cite{sandars68}.
By considering these interactions into account, the total atomic Hamiltonian 
 can be written as,
\begin{eqnarray}
H\, =\, H_0\, + \, H_{EDM}
\label{totalH}
\end{eqnarray}
where $H_0$ is the unperturbed Hamiltonian given by Eq.(\ref{H0}) and  
$H_{EDM}$ is the part of the Hamiltonian perturbed either by intrinsic e-EDM contribution or by the S-PS EDM contribution and they are given respectively by,
\begin{eqnarray}
H_{EDM}^{e}\, = \, 2\, i\, c\, d_e \sum_j \beta_j\, \gamma_j^5 \, \vec {p}_j\,^2
\end{eqnarray}
and
\begin{eqnarray}
H_{EDM}^{s-ps}\, =\, \frac{i G_F}{\sqrt{2}}C_S A \sum_j \beta_j\, \gamma_j^5\, \rho_\mathcal{N}(r_j)
\end{eqnarray}
where $d_e$ is the coupling constant for e-EDM, $\gamma^5$ is the product of the Dirac matrices i.e, $\gamma^5 = i \gamma^1 \gamma^2  \gamma^3 \gamma^4 $, $\,\vec p_j$ is the momentum vector of the $j^{th}$ electron, 
$G_F$ is the Fermi coupling constant, $C_S$ is the dimensionless S-PS constant, $A$ is the mass number of the atom and $\rho_\mathcal{N}$ is the nuclear density. The $H_{EDM}$ mixes the atomic states of opposite parities but
 with the same angular momentum. As its strength is sufficiently weak, we
consider only up to the first-order perturbation in wavefunction. Thus, the modified atomic wavefunction for the valence electron state '$v$' is given by,
\begin{eqnarray}
|\Psi_v \rangle \cong |\Psi_v^{(0)} \rangle + \lambda\, |\Psi_v^{(1)} \rangle
\end{eqnarray}
where $\lambda = d_e$ for e-EDM, whereas, $\lambda = G_F C_S A /\sqrt{2}$ for S-PS EDM. 

In the RCC ansatz, the cluster operators for calculating the perturbed wavefunctions are given by,
\begin{eqnarray}
T = T^{(0)} + \lambda\, T^{(1)} ; ~\, \text{and} ~\,  S_v =  S_v^{(0)} + \lambda \,S_v^{(1)}
\end{eqnarray}
where $T^{(1)}$ and $S_v^{(1)}$ are the first-order corrections to the unperturbed cluster operators  $T^{(0)}$ and $S_v^{(0)}$, respectively. In the relativistic CCSD approximation, the perturbed and unperturbed cluster operators are taken to be, $T = T_1 + T_2$ and $S_v = S_{v_1} + S_{v_2}$. 
 We have excited all the core electrons to all possible virtual states in the present calculations. To obtain better accuracies of the wavefunctions we have calculated the perturbed and unperturbed cluster amplitudes with a convergence of less than $10^{-6}$ in all the cases.

The expectation value of an atomic EDM ($D_a$) for the state $|\Psi_v \rangle$ is given by,
\begin{eqnarray}
\label{da}
D_a \cong \frac {\langle \Psi_v|\, \vec{D}\, |\Psi_v \rangle } {\langle \Psi_v^{(0)}|\Psi_v^{(0)} \rangle}
\end{eqnarray}
where $\vec {D} = e \vec{r}$ is the electric dipole moment operator. 
The final expression for the EDM enhancement factor ($R = \frac{D_a}{d_e}$; or $S= \frac{D_a}{G_F C_S A /\sqrt{2}}$) in terms of the perturbed and unperturbed cluster amplitudes is given by,
\begin{widetext}
\begin{eqnarray}
R \,(\text{or}\, S) =  {\displaystyle \frac{\langle \Phi_v |\left\{ \{1+ S_v^{(0)^\dagger}\} \overline{D^{(0)}} \{ T^{(1)} + T^{(1)} S_v^{(0)} + S_v^{(1)}\} + \{ S_v^{(1)^{\dagger}} + S_v^{(0)^{\dagger}} T^{(1)^{\dagger}} + T^{(1)^{\dagger}}\} \overline{D^{(0)}} \{1+ S_v^{(0)}\} \right\} |\Phi_v \rangle }{\langle \Phi_v |\, e^{T^{(0)^\dagger}}\, e^{T^{(0)}} \,+\, S_v^{(0)^{\dagger}}\, e^{T^{(0)^{\dagger}}}\, e^{T^{(0)}}S_v^{(0)}\, | \Phi_v \rangle}}
\end{eqnarray}
\end{widetext}
where $\overline{D^{(0)}} = e^{{T^{(0)}}^\dagger}\, \vec{D}\, e^{T^{(0)}}$. 
In the above expression the DF term is contained in $\overline{D^{(0)}}\, S_v^{(1)}$. It corresponds to considering $\vec{D}$ in $\overline{D^{(0)}}$ and one order of EDM interaction and zero order of residual Coulomb interaction in $S_v^{(1)}$. It can be explicitly written in the DF approximation as,
\begin{eqnarray}
\label{dfedm}
 D_a ^{(DF)} \, = \,2\, \sum_{I \neq v} \frac{\langle \Phi_v \vert H_{EDM} \vert \Phi_I \rangle \langle \Phi_I \vert \vec{D} \vert \Phi_v \rangle}{E_v - E_I}
\end{eqnarray}
where the intermediate states $\vert \Phi_I \rangle $ differs from the valence reference state $\vert \Phi_v \rangle$ which is $5s$ for Rb and $6s$ for Cs either by a single valence or core orbital and of opposite parity. 

It has to be noted that the accuracy of the coupled-cluster results depend on the size of the configuration space of the RCC wavefunctions considered. However, an increase in the configuration space leads to a larger  memory requirement and also computational time. Thus, one is limited by the maximum size of the configuration space that one has considered in an actual calculation. Our preliminary results with a medium sized basis set and with a limited number of multipoles satisfying Coulomb interaction selection rules were published before in \cite{nataraj1}. However, here we have taken more than 5 different sets of large configuration spaces with all possible multipoles and  systematically studied the nature of the correlation trends of the dominant RCC terms in Eq.(\ref{dfedm}) such as, $\overline{D^{(0)}}\, S_v^{(1)} $, $\overline{D^{(0)}}\, T^{(1)} $, and $S_v^{(0)^\dagger}\, \overline{D^{(0)}}\, S_v^{(1)} $ with their hermitian conjugates contributing to the e-EDM enhancement factor ($R$) for the Rb atom and plotted the variation in the magnitude of the correlation contribution of these terms as (a), (b) and (c) in Figure 1, respectively. 
The RCC formulation contains all-order relativistic Coulomb correlation effects like core-correlation through the term $\overline{D^{(0)}}\, T^{(1)} $, pair-correlation and core-polarization through the term $\overline{D^{(0)}}\, S_v^{(1)} $. The difference between $D S_{v_1}^{(1)}$ and the DF terms is the largest of all the correlation contributions.
It is clear from all the above three diagrams that the magnitude of correlation increases with the increase in the size of the basis set and more or less gets saturated beyond $95$. It is interesting to observe 
the same convergence trend in the e-EDM enhancement factor $R$ shown in the last diagram (d) with the size of the basis set. 
Thus, we obtain the convergence in the magnitude of EDM enhancement factor by taking sufficiently large  number of basis functions to represent the space spanned by the RCC wavefunctions.
Using the same basis sets we also obtain the EDM enhancement factors due to S-PS electron-nucleus interactions ($S$) for Rb. We have computed the EDM enhancement factor $R$ for a few basis sets of size larger than $100$ for Cs. The value of $S$ for Cs is taken from \cite{bijaya-sps}.
Finally the converged results both for Rb and Cs  along with the Dirac-Coulomb results are tabulated in Table I. We have also compared our results with some of the published ab-initio methods based on finite order MBPT \cite{johnson86,hartley90}, MCDF+MBPT \cite{bpdas94,bpdas} and semi-empirical one-electron theories \cite{sandars66}. 
\begin{figure}
\label{corrplots}
\caption{The variation in the trends of the dominant RCC terms such as, (a) $\overline{D^{(0)}}\, S_v^{(1)}\, $ (b) $\overline{D^{(0)}}\, T^{(1)} $ (c) $S_v^{(0)^\dagger}\, \overline{D^{(0)}}\, S_v^{(1)} $ and (d) Total e-EDM enhancement factor ($R$) with the size of the basis set for Rb.}
\includegraphics{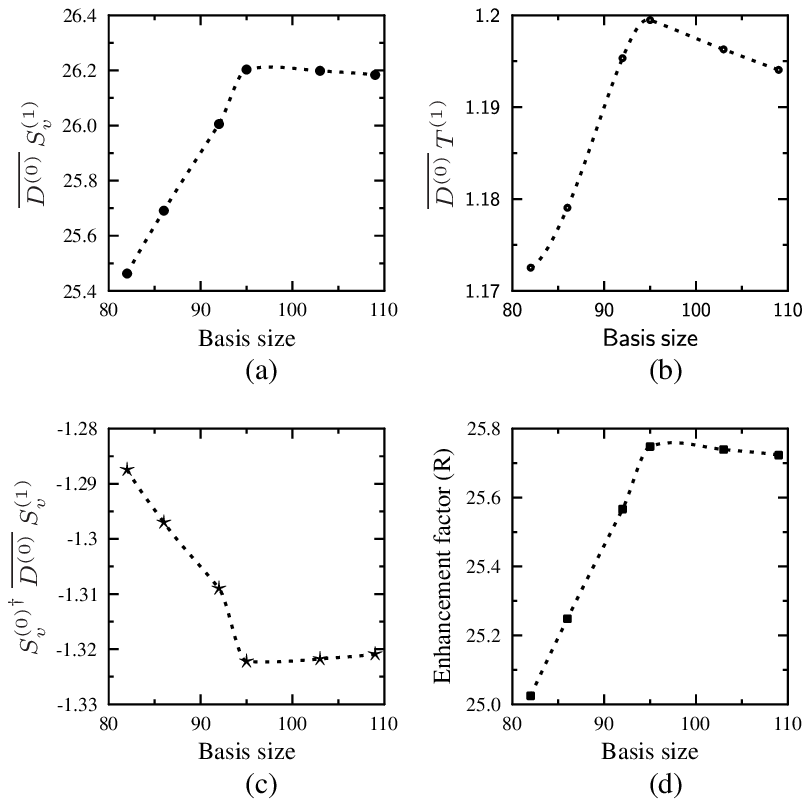}
\end{figure}

\begin{table}
\label{edmresults}
\caption{The EDM enhancement factors of the ground state of Rb and Cs atoms. R $( = \frac{D_a}{d_e} )$ and S $( = \frac{D_a}{G_F C_S A /\sqrt{2}} )$ are the enhancement factors due to electron EDM and scalar--pseudo-scalar electron-nucleus EDM respectively.} 
\begin{ruledtabular}
\begin{tabular}[t]{cccccc}
Atom &  \multicolumn{2}{c}{R} & \multicolumn{2}{c}{S} & Reference\\ \cline{2-3} \cline{4-5}
   &     DF &  Total & DF & Total & \\
\hline \\
Rb & 19.55 & 25.74 & 11.91 &  15.64 & This work\\
   &       & 24.6  &       &        & \cite{johnson86} \\
   &       & 25.68 &       &  16.47 & \cite{bpdas94}\\
   &       & 24.0      &    &     & \cite{sandars66} \\
Cs & 94.19 & 120.53 &  &  & This work \\
   &  &  & 52.34 & 72.44  & \cite{bijaya-sps} \\
   &       & 114.9  &  & & \cite{johnson86}\\
   &       & 114    &  & & \cite{hartley90}\\
   &       & 130.5  &       &       &  \cite{bpdas}\\
   &       & 119    &  & & \cite{sandars66} \\
\end{tabular}
\end{ruledtabular}
\end{table}

The EDM matrix element is sensitive to the wavefunction near the nuclear region where as the dipole matrix element is sensitive to the wavefunction at far-off regions from the nucleus. Thus there is a greater need for more reliable relativistic calculations of the atomic wavefunctions 
 of both the nearby and far-off states, from the nucleus. The RCC theory employed in the present work has such a potential. The accuracy of EDM results will depend on; $(i)$ the excitation energies, as the reliable values of these are crucial as the EDM enhancement factor inversely depends on the energy differences as shown in Eq.(\ref{dfedm}), $(ii)$ the spontaneous electric dipole (E1) transition amplitude for the transition from the lowest $^2P_{1/2}$ state to the ground state and $(iii)$ the magnetic dipole hyperfine constants (A) of the lowest $^2S_{1/2}$ and $^2P_{1/2}$ states. To a good approximation, the error in the EDM matrix elements can be taken as $\approx \sqrt{A_{\,^2S_{1/2}} \cdot \, A_{\,^2P_{1/2}}}$. 
 We have shown all the above discussed property results for Rb in Table II and compared them with those of the published experimental results some of which are known up to fairly high accuracies such as, for example, Rb ground state hyperfine energies are considered for the frequency standards and our results agree reasonably well with them. 

\begin{table}
\label{hypresults}
\caption{The excitation energy (in $cm^{-1}$) and the E1 transition matrix element (in $a.u.$ ) for $5\,^2S \rightarrow 5\,^2P_{1/2}$ transition and the magnetic dipole hyperfine constants of the  $5\,^2S_{1/2}$ and $5\,^2P_{1/2}$ states of Rb atom ($^{85}Rb ;\, I^{\pi}=5/2\,^-$).} 
\begin{ruledtabular}
\begin{tabular}[t]{lccc}

Property & Transition/ & \multicolumn{2}{c}{Magnitude} \\ \cline{3-4}
         & State       &  This work & Expt. [Ref.] \\ 
\hline \\
Excitation energy      & $5\,^2S \rightarrow 5\,^2P_{1/2}$ & 12579.87 &  12578.95 \cite{nist1}  \\
E1 tr. amplitude      &  $5\,^2S \rightarrow 5\,^2P_{1/2}$& 4.26 &  4.23 \cite{volz96} \\
Hyperfine Constant  & $5\,^2S_{1/2}$         &  1009.33 & 1011.91 \cite{vanier74}   \\
Hyperfine Constant      & $5\,^2P_{1/2}$ &  117.71  &  120.64 \cite{vasant06}\\ 
\end{tabular}
\end{ruledtabular}
\end{table}
The differences between the computed and the accurately known experimental results presented in Table II, give the individual errors and by adding the errors in quadrature we estimate that the maximum error in our EDM results could be around $ 0.7 \%$. We have also obtained the Cs results of similar quality. 

The limit on the e-EDM will be obtained by combining the precise values of both the atomic EDM measurements and the theoretical enhancement factors. The better e-EDM limit from Cs EDM experiments has resulted from \cite{sudha89}. 
By combining their experimental result of Cs EDM and our computed enhancement factor we obtain the limit $ (-1.5 \pm 3.2) \times 10^{-26} e-cm$. However, the best ever e-EDM limit is obtained from Tl EDM experiment as $ (6.9 \pm 7.4) \times 10^{-28} e-cm$ by Regan et al. \cite{regan02}. There are a few state-of-art EDM experiments which are currently underway to measure atomic EDMs in Rb and Cs \cite{weiss, gould, heinzen} aiming to achieve almost 2-3 orders of higher sensitivity than the reach of the current experimental setups and when these attempts come to fruition, by using our enhancement factors a better limit on intrinsic e-EDM can be achieved.

In conclusion, we have calculated the EDM enhancement factors of the ground states of the paramagnetic atoms such as, Rb and Cs atoms due to both the intrinsic electron EDM and the scalar--pseudo-scalar electron-nucleus interactions using the RCC theory which inherently has an all-order relativistic many-body nature. We have systematically studied different correlation terms, which contribute dominantly to atomic EDM, for a number of basis sets of varying configuration spaces. We have estimated the conservative upper limit of the error bar in our EDM results to be $1 \%$ by studying various other atomic properties of the lowest $^2S_{1/2}$ and $^2P_{1/2}$ states of Rb and Cs which dominantly contribute to the EDM enhancement factors of their corresponding ground states and comparing them with the accurately known experimental results. 

We acknowledge Dr.Rajat Kumar Chaudhuri of Indian Institute of Astrophysics, Bangalore, India for developing certain parts of the codes. We also thank the staff of CDAC Terra-flopp Supercomputing Facility at Bangalore, India for their support and the computational facility for carrying out this work.


\begin{thebibliography}{}
\bibitem{regan02}
B. C. Regan, E. D. Commins, C. J. Schmidt and D. DeMille, Phys. Rev. Lett. {\bf 88}, 071805 (2002)
\bibitem{weiss}
D. S. Weiss, F. Fang and J. Chen, Bull. Am. Phys. Soc. APR03, J1.008 (2003)
\bibitem{gould}
J. M. Amini, C. T. Munger, Jr., and H. Gould, Phys. Rev. A {\bf 75}, 063416 (2007) 
\bibitem{heinzen}
Kittle, M.; Burton, T.; Feeney, L.; Heinzen, D. J.,\\ Bibliographic Code: 2004APS..DMP.P1056K
\bibitem{nasa-edm}
Electronic Address: \\
http://funphysics.jpl.nasa.gov/technical/lcap/edm-x.html
\bibitem{pospelov}
For reviews, see Maxim Pospelov and Adam Ritz, Annals of Physics, {\bf 318}, 119 (2005); 
J. S. M. Ginges and V. V. Flambaum, Phys. Rep. {\bf 397}, 63 (2004)
\bibitem{huber07}
Stephen J. Huber, Maxim Pospelov and Adam Ritz, Phys. Rev. D {\bf 75}, 036006 (2007); 
A. M. Kazarian, S. V. Kuzmin and M. E. Shaposhnikov, Phys. Lett. B {\bf 276}, 131 (1992) 
\bibitem{bernreuther91}
Werner Bernreuther and Mahiko Suzuki, Reviews of Modern Physics {\bf 63}, 313 (1991)
\bibitem{bernreuther02}
Werner Bernreuther, Lecture notes in Physics, {\bf 591}, 237 (2002), (Springer Berlin/Heidelberg)
\bibitem{bijaya-pnc}
B. K. Sahoo, Rajat Chaudhuri, B. P. Das and D. Mukherjee, Phys. Rev. Lett. {\bf 96}, 163003 (2006)
\bibitem {nataraj1}
H. S. Nataraj, B. K. Sahoo, B. P. Das, R. K. Chaudhuri and D. Mukherjee, J. Phys. Conf. Sr., {\bf 80}, 012050, (2007)
\bibitem{bijaya-sps}
B. K. Sahoo, R. K. Chaudhuri, B. P. Das, D. Mukherjee and E. P. Venugopal, arXiv:physics/0509070
\bibitem{johnson86}
W. R. Johnson, D. S. Guo, M. Idres and J. Sapirstein, 
Phys. Rev. A {\bf 34}, 1043 (1986)
\bibitem{hartley90}
A. C. Hartley, E. Lindorth and A. M. Martensson-Pendril, J. Phys. B. {\bf 23}, 3417 (1990)
\bibitem{bpdas94}
Alok Shukla, B. P. Das and J. Andriessen, Phys. Rev. A {\bf 50}, 1155 (1994)
\bibitem{bpdas}
B. P. Das, Lecture Notes in Chemistry {\bf 50}, 411 (1988) 
\bibitem{sandars66}
P. G. H. Sandars, Phys. Lett. {\bf 14}, 194 (1965); Phys. Lett. {\bf 22}, 290 (1966)
\bibitem{sandars68}
P. G. H. Sandars, J. Phys. B {\bf 1}, 511 (1968)
\bibitem{nist1}
Ralchenko, Yu., Kramida, A.E., Reader, J., and NIST ASD Team (2008).
\bibitem{volz96}
U. Volz and H. Schmoranzer, Phys. Scripta, {\bf T65}, 48 (1996)
\bibitem{vanier74}
J. Vanier, J. F. Stimard and J. S. Boulanger, Phys. Rev. A {\bf 9}, 1031 (1974)
\bibitem{vasant06}
D. Das and V. Natarajan, Eur. Phys. J. D {\bf 37}, 313 (2006)
\bibitem{sudha89}
S. A. Murthy, D. Krause, Jr., Z. L. Li and L. R. Hunter, Phys. Rev. Lett. {\bf 63}, 965 (1989)

\end{thebibliography}
\end{document}